**Buffon Needle Problem Application to Space Exploration**

**Sedelnikov A.V.**

**Samara state aerospace university**

In this article the possibility of application of classical Buffon needle problem to the investigation of orientation engine firing problem has been investigated. Such an approach makes it possible to get a reliable EP of this undesired event without using a more complicated analysis.

When gravitation-sensible operations are carried out in-space it is important to provide all necessary conditions for their successful carrying out [1,2]. One of those conditions is the orientation engine non-firing during the operation. Otherwise the carrying out the experiment shell fail.

As it happens, it is possible to anticipate the engine firing probability during such an operation using the classical Buffon needle problem [3]. Let us specify how to do it.

Let us assume that the test duration time run $\tau$ shorter then the average period between the engine firings $t_{cp}$ . Let us draw on the plane a vertical time axis where there are shown the rocket engine firings moments $t_1 < t_2 < t_3 < ... < t_m$, where "m" means the total number of engine firings during the space vehicle flight time summary. Then let's draw the parallel lines through the shown points, these parallel lines shall bee perpendicular to the axes, though we get the grid with parallel lines plane like the way it is provided in the Buffon problem. In order to make the next true reasoning let us assume that that the intervals between the engine firings are equal because the distances between lines in the Buffon problem are constant.

Let's assume that during the spice vehicle flight "n" experiments are made, the durations of such experiments are different and is equal $\tau_1, \tau_2, ..., \tau_n$ accordingly. At that, the start time of each next test is random and doesn't depend on the next and previous tests. The durations $\tau_i$ , where $i = 1, 2, \ldots, n$ are not interdependent. In this case the values $\tau_i$ correspond with the needle projections to the direction which is perpendicular to the lines. In the Buffon problem one of the chance variables shall be the fallen needle angle of slop towards the line directions, another chance variable shall be the distance from its underside end to the nearest line above [3]. In the specified task the first chance variable shall be $\tau_i$, which is corresponded with the product of the needle length and the angle of slop sine. The second chance variable shall be the interval between the start of the *i-st* experiment and the next engine firing. So each experiment test could be regarded as throw of needle. At that, the situation shall not be changed depending on the way of the throw of needle: in series (when the tests are made one by one at time intervals) or some needles are thrown at once (a test series is made at the same time or if the test moments superimpose on one another).

The engine firing during the test in this case looks like a needle crossing one of the lines. Let's assume that before the $S-st$ test start the engine fired $k$ times, than, if we assume that $T_s$ shall be the start moment of the $S-st$ test, we shall get the condition of the above specified event – the engine firing during the test:

$$T_s - k \cdot t_{cp} \leq \tau_s ,$$

which shall be correspond with the condition specified in [3, page 12]. Than the event probability shall be

$$P(A) = \frac{2\tau}{\pi t_{cp}} \tag{1}$$

So the classical tasks of the theory of probability could be successfully used by solution of real tasks in the field of space explorations. At that, the algebraic expression (1) corresponds with the experimental data of the experiment without operating special actions to avoid the engine firing during the operating. As an example could be taken in the American space station "Skylab" where about 300 series of different experiments and operations has been made and carried out.

Address for connection: Russia, state Samara, 443026, p.b. 1253, Dr. Andry Sedelnikov

Address on Russian: Россия, г. Самара, а/я 1253.

E\mail: axe_backdraft@inbox.ru